\documentclass[twocolumn,superscriptaddress]{revtex4-1}
\usepackage{graphicx,amsmath}
\usepackage[dvipsnames]{xcolor}
\usepackage{subcaption}
\usepackage{enumerate}
\usepackage{float}
\usepackage{color}
\usepackage{xcolor}
\pagecolor{white}
\pdfoutput=1
\usepackage{calrsfs}

\newcommand{\hbd}{\hat{b}^+}
\newcommand{\hb}{\hat{b}}

\newcommand{\ket}[1]{\left| #1 \right\rangle}

\newcommand{\expect}[1]{\left\langle#1\right\rangle}
\newcommand{\eea}{\end{eqnarray}}
\newcommand{\bea}{\begin{eqnarray}}
\newcommand{\ee}{\end{equation}}
\newcommand{\be}{\begin{equation}}

\newcommand{\sch}{Schr\"{o}dinger }

\begin{document}
\title{How to describe collective decay of uncoupled modes in the input-output formalism}

\author{Tzula B. Propp}
\affiliation{Center for Quantum Information \& Control, University of New Mexico, Albuquerque, NM 87131}

\begin{abstract}
We extend the input-output formalism to study the behavior of uncoupled discrete modes (bosonic cavity modes and fermionic qubits) when they decay to the same Markovian continuum. When the continuum interacts with only a single mode, this decay is irreversible. However, when multiple modes decay to the same Markovian continuum they develop correlations and decay collectively. In the input-output formalism these correlations manifest in additional terms in the quantum Langevin equation. For two modes this collective decay can dramatically extend the lifetimes of both modes (Dicke subradiance) and, within the single-mode subsystem, induces non-Markovian memory effects including energy back flow. 

\end{abstract}

\maketitle

\section{Introduction}

Naively, the Markov approximation prevents both information and energy back flow into a system. While true for a system as a whole, this can breakdown for individual subsystems. In the absence of a  coupling term in a system Hamiltonian, different modes in an open quantum system will still develop strong correlations if they radiate into the same quantum field--the electromagnetic continuum of states \cite{dicke54}. Collective decay, including both Dicke sub-radiance and superradiance, have been observed experimentally in atomic ensembles via near field interactions \cite{Gross19823,Kalachev2006,kalachev2007,kimble2015,Cidrim2020} and far field interactions \cite{vanenk1999,solano2017,atomwire2017,manzoni2018,sinha2020,masson2020,massonmany2020} (e.g. a bi-directional transmission channel such as an optical fiber, allowing for a propagating mediating field), and can give rise to non-Markovian dynamics \cite{sinha2020non}. Even without a bi-directional near or far-field interaction initially uncoupled degrees of freedom in a system can spontaneously radiatively couple \cite{cardimona1983,tscherbul2014}, generating coherence between system degrees of freedom \cite{altenmuller1995}. 

In this paper, we derive an extension to the input-output formalism \cite{input1985} for open quantum systems with multiple modes radiating into the same continuum such that they decay collectively. Building upon early work in quantum network theory \cite{yurke1984}, input-output theory is a powerful theoretical toolkit where quantum devices are modeled as networks of modes with quantized inputs and outputs corresponding to continuum degrees of freedom \cite{input1985} with extensions to fermionic systems \cite{bi1999,search2002,gardiner2004} and applications in squeezed-mode generation and parametric amplification \cite{Collett1984,kimble1986}, cascaded quantum systems \cite{carmichae1993,gardiner1993}, superconducting circuits \cite{Wilson2011,Korotkov2011}, finite-temperature optomechanics \cite{barchielli2015}, single photon detectors \cite{young2018,ProppNet}, general quantum amplifiers \cite{caves1982,proppamp,combes21}, quantum transduction \cite{vanEnk2022}, and a wide range topics in quantum optics (for a thorough review see Ref.~\cite{combes2017})\footnote{The input-output formalism is also in the same spirit as the LSZ reduction, paving the way to natural connections to high-energy physics .}. While adding more continua is straightforward (the effect is generally additive \cite{giacosa2020}), adding additional system degrees of freedom coupled to the same continua requires modification of the standard quantum Langevin equation.

 \begin{figure}[t] 
	\includegraphics[width=1\linewidth]{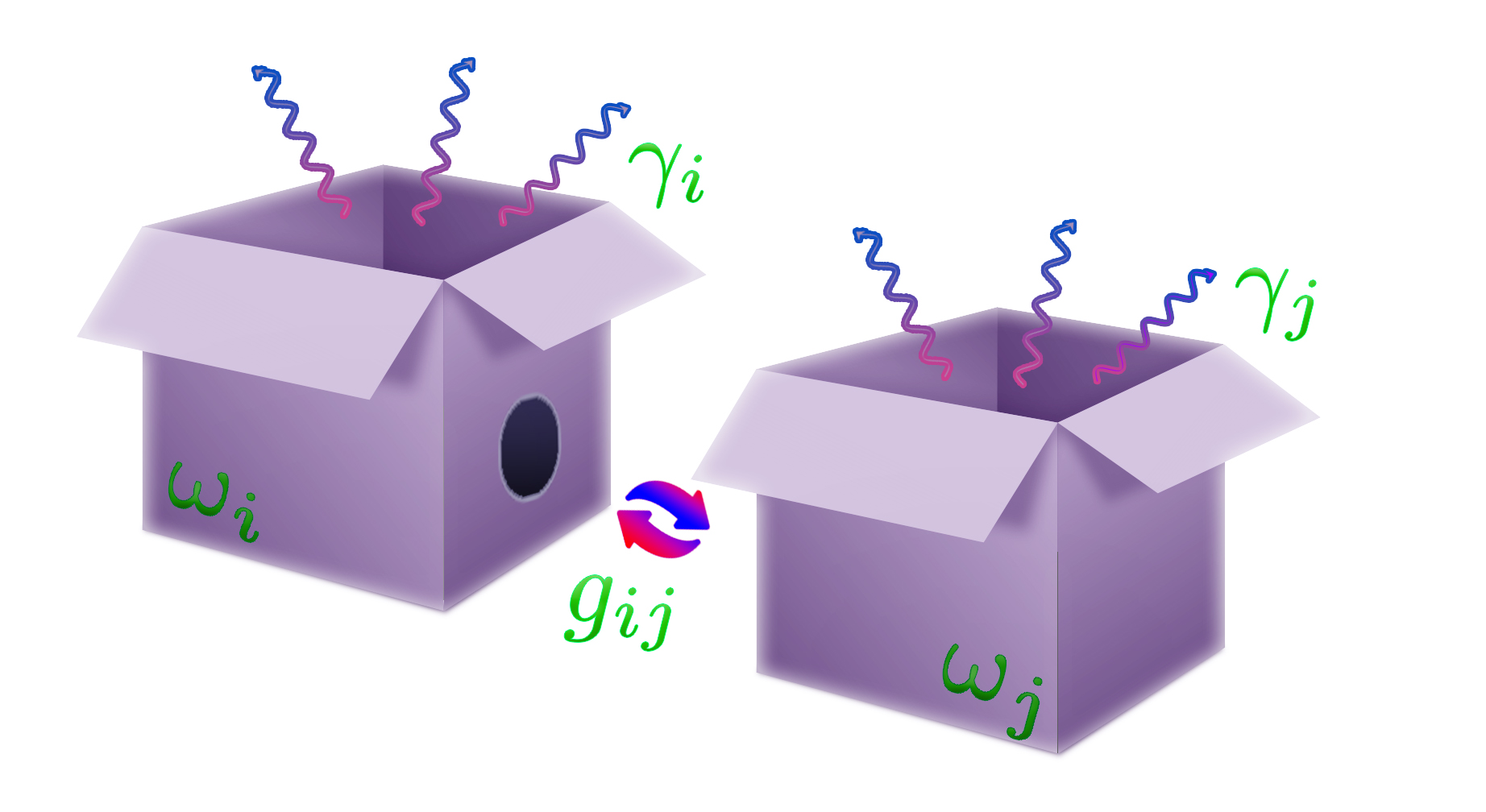}
	\caption[]{Two discrete modes (depicted as boxes) with resonance frequencies $\omega_i$ and $\omega_j$ emit light into a shared global continuum of states at rates $\gamma_i$ and $\gamma_j$. For generality, we also allow them to couple to each other (directly or indirectly) at a rate $g_{ij}$, which we will later set to zero. In the Heisenberg picture, each mode is described by a time-dependent discrete mode operator ($\hat{c}_i$ and $\hat{c}_j$) and the shared continuum is described by the time-dependent continuum mode operator $\hat{b}(\omega)$.}
	\label{networkschem}
\end{figure}

The main result of our paper is the generalized quantum Langevin equation 

\bea\label{quantumlangevin}
\dot{\hat{c}}_i(t)&=& -i\omega_i \hat{c}_i(t)-i\!\sum\limits_{j} g_{ij} \hat{c}_j(t) -\! \sum_j \frac{\sqrt{\gamma_i\gamma_j}}{2} \hat{c}_j(t) - \sqrt{\gamma_i} \hat{b}_{\rm in}(t)\nonumber \\
\eea describing the Heisenberg-picture evolution of a collection of bosonic system mode annihilation operators $\hat{c}_i(t)$ with mode resonance frequencies $\omega_i$ and decay rates $\gamma_i$ to a shared continuum described by a single noise operator $ \hat{b}_{\rm in}(t)$ (Fig. \ref{networkschem}). For generality, we have also included inter-mode couplings $g_{ij}$ which are assumed w.l.o.g. to be real. As we will see from (\ref{quantumlangevin}), it makes a great deal of difference whether a set of modes decay locally (each to their own continuum) or globally (to a single, shared continuum). We will also contrast (\ref{quantumlangevin}) with the standard quantum Langevin equation, where (naively) applying the results of Ref.~\cite{input1985} yields

\bea\label{langevinwrong}
\dot{\hat{c}}_i(t)&=& -i\omega_i \hat{c}_i(t)-i\!\sum\limits_{j\neq i} g_{ij} \hat{c}_j(t) -\frac{\gamma_i}{2} \hat{c}_j(t) - \sqrt{\gamma_i} \hat{b}_{\rm in}(t)\nonumber\\
\eea which is missing the correction terms due to spontaneous radiative coupling. For systems where multiple bosonic modes decay to a single continua, (\ref{quantumlangevin}) and (\ref{langevinwrong}) are not equivalent. Examples include multiple modes in optical cavities \cite{Fan2004,Houck2015,Valetin2022}, structured continua with multiple pseudomodes \cite{Hughes2018}, quantum dots \cite{Livache2019}, multimode laser dynamics \cite{bandelow2014}, and, for fermions, multiple P-orbitals in an atom or molecule \cite{Higginbottom2016}. Crucially, we are \emph{not} studying cascaded quantum systems where one quantum systems drives another in series \cite{carmichae1993}; for our purposes, we are considering multiple modes decaying in parallel.


In addition to descriptions of multiple modes decaying to the same continuum using other formalisms (dressed states \cite{cardimona1983,tscherbul2014}, collision models \cite{Ciccarello2017}, and multimode cavity QED \cite{Houck2015, Koll2015,schuster2015}), there have been a few descriptions of systems with spontaneous radiative coupling in the input-output formalism. In studying fermionic systems, Search et al. \cite{search2002} include the correct modification to the quantum Langevin equation, and (correctly) interpret it as a indirect coupling between modes induced by the shared continuum of states (that is, spontaneous radiative coupling). Additionally, an equation similar to the generalized quantum Langevin equation (\ref{quantumlangevin}) appears in the appendix of Ref.~\cite{xiao2020} without derivation or discussion. Therefore, the primary aim of this paper is to carefully derive and discuss the form of the generalized quantum Langevin equation and the physics contained therein. We organize the paper as follows. In section II, we rederive the input-output theory for multi-mode systems decaying to a shared global continuum, arriving at a modified quantum Langevin equation (\ref{quantumlangevin}). We then study uncoupled parallel networks, emphasizing the strong correlations between discrete mode amplitudes. In section III, we further specialize to two two-mode examples: a) two fermionic qubits and b) two bosonic cavity modes (which can be used to make continuous variable qubits \cite{munro2000}). For both cases, we observe collective decay phenomena including whole system subradiance and subsystem population revival, the latter of which we explore in more detail in section IV. Finally in section V, we propose applications for reservoir engineering in quantum information experiments and for modeling of non-Markovian open quantum systems in the style of the method of pseudomodes \cite{pseudomodes2,pseudomodes3,pseudomodes4,pseudomodes1,pleasance2021}, before briefly revisiting general interpretation in the conclusions.


\section{Theory}

We now carefully consider the case of a quantum network of multiple discrete modes decaying to a single bosonic continuum (Fig. \ref{networkschem}), following closely (but not identically) along the lines of Gardiner and Collet's seminal paper \cite{input1985}. Crucially, we will allow independent system operators to couple to the same Markovian continuum in parallel. We work in natural units ($\hbar = 1$) and in $1$D, noting that the results of $1$D input-output theory hold in $3$D \cite{gardiner1986,vanenk2004}.

Discrete modes are operators with sharply-peaked spectra in the Heisenberg picture. The Hamiltonian describing a network of arbitrary discrete modes interacting with each other as well as a single bosonic continuum can be divided into three terms $\hat{H} = \hat{H}_{\rm sys} +  \hat{H}_{\rm b}+ \hat{H}_{\rm int}$.  The first term is a system Hamiltonian 

\bea\label{Hsys}
\hat{H}_{\rm sys} &=& \sum_i \omega_i \hat{c}^\dagger_i\hat{c}_i + \sum\limits_{\{i,j\}} g_{ij}\left(\hat{c}^\dagger_i\hat{c}_j+\hat{c}^\dagger_j\hat{c}_i\right)
\eea where $\omega_i$ is the resonance frequency of the $i$th discrete mode, $g_{ij}$ is the coupling between the $i$th and $j$th discrete mode (w.l.o.g. assumed to be real), and $\hat{c}_i$ and $\hat{c}^\dagger_i$ are the lowering and raising operators for the $i$th discrete mode. At this stage, the system discrete mode operators $\hat{c}_i$  are arbitrary and may satisfy bosonic, fermionic, or anyonic commutation relations. Similarly, we leave the couplings $g_{ij}$ arbitrary to incorporate a general network geometry, except to define $g_{ii}=0\,\forall i$. (Any self-interaction can be absorbed into the discrete mode energy $\omega_i$.)

The second term is a bosonic Hamiltonian
\bea\label{Hb}
\hat{H}_{\rm b} &=& \int d\omega \omega \hat{b}^\dagger(\omega) \hat{b}(\omega)
\eea where $ \hat{b}(\omega)$ and $ \hat{b}^\dagger(\omega)$ are the annihilation and creation operators for the monochromatic electromagnetic mode with frequency $\omega$ within the bosonic $1$D continuum of modes, satisfying the canonical commutation relation $[\hat{b}(\omega),\hat{b}^\dagger (\omega')] = \delta(\omega-\omega')$. 

The third term is the interaction Hamiltonian

\bea\label{Hint}
\hat{H}_{\rm int} &=& i \int d\omega \sum_i \kappa_i(\omega)\left(\hat{b}^\dagger(\omega)\hat{c}_i-\hat{c}^\dagger_i  \hat{b}(\omega)\right)
\eea where $\kappa_i(\omega)$ is a frequency-dependent coupling describing exchange of excitations between the discrete mode network and the continuum at a single location $x=0$ in the rotating wave approximation (which is valid in the weak-coupling limit). If one relaxes the assumption that the discrete modes occupy the same position, there are additional factors $e^{x_i \omega_i/c}$ in the interaction Hamiltonian but the derivation below remains effectively unchanged. 

We now move to the Heisenberg picture. Here the equations of motion describing continuum and discrete mode time-dependent operator evolution are

\bea\label{Hmot}
\dot{\hat{b}}(t,\omega) &=& -i\omega \hat{b}(t,\omega) + \sum_j \kappa_j(\omega) \hat{c}_j (t)\\
\dot{\hat{c}}_i(t)&=& -i[\hat{c}_i(t),\hat{H}_{\rm sys}]- \int d\omega \kappa_i(\omega) [\hat{c}_i(t),\hat{c}^\dagger_i(t)]\hat{b}(t,\omega)\nonumber \\
&\rightarrow& -i\omega \hat{c}_i(t)-i\sum_{j} g_{ij} \hat{c}_j (t)- \int d\omega \kappa_i(\omega) \hat{b}(t,\omega)\nonumber
\eea where in the last line we have assumed the discrete mode operator $\hat{c}_i(t)$ to be bosonic and satisfy the canonical bosonic commutation relation for discrete modes $[\hat{c}_i(t),\hat{c}^\dagger_i(t)]=1$. This will also describe fermionic (and anyonic) discrete modes in the weak-field limit so that there is (at most) a single excitation in each mode. Detailed study of strong-field fermionic discrete modes are beyond the scope of the present paper, but we give an analogous derivation of collective decay for fermionic modes in the Appendix A.

The first line of (\ref{Hmot}) has the formal solution

\bea\label{formsol}
\hat{b}(t,\omega) &=& e^{-i\omega(t-t_0)} \hat{b}(t_0,\omega)+\int_{t-t_0}^t dt' e^{-i\omega (t-t')} \sum_j \kappa_j(\omega) \hat{c}_j(t') \nonumber \\
\eea where $\hat{b}(t_0,\omega)$ describes initial value of the continuum field at the time $t_0$. Inserting (\ref{formsol}) into the second line of (\ref{Hmot}) yields an equation of motion for discrete mode evolution with the continuum field integrated out

\begin{widetext}\bea\label{modesol}
\dot{\hat{c}}_i(t)&=& -i\omega_i \hat{c}_i(t)-i\sum\limits_{j} g_{ij} \hat{c}_j(t) - \int d\omega  \kappa_i(\omega)e^{-i\omega(t-t_0)} \hat{b}(t_0,\omega)\\
&-& \int d\omega \kappa_i(\omega)\int_{t-t_0}^t dt' e^{-i\omega (t-t')} \sum_j \kappa_j(\omega) \hat{c}_j(t') . \nonumber
\eea \end{widetext}

We now invoke the first Markov approximation $\kappa_i(\omega) = \sqrt{\frac{\gamma_i}{2\pi}}$ for each discrete mode with one key modification; in addition to assuming the decay rates $\gamma_i$ to be constant over each discrete mode's frequency $\omega_i$, we must also assume that they are \emph{also} constant over each \emph{other} discrete mode's frequency $\omega_j$ in order to pull them out of the frequency integral in (\ref{modesol}). To be clear, this is a slightly stronger assumption than is made in Ref.~\cite{input1985}. However, it is subsumed by the standard Markov approximation when the energy spacing $\omega_i-\omega_j$ is of the same order as the decay rates $\gamma_i$ (that is, the spectral half-widths of the discrete modes).

We also define an input field 

\bea\label{bin}
\hat{b}_{\rm in}(t) &=& \frac{1}{\sqrt{2\pi}} \int d\omega e^{-i\omega(t-t_0)} \hat{b}(t_0,\omega)
\eea satisfying the canonical commutation relation $[\hat{b}_{\rm in}(t),\hat{b}_{\rm in}^\dagger (t')] = \delta(t-t')$, and make the usual arguments for 

\bea\label{delta}
\int_{t-t_0}^t dt' f(t') \delta(t-t') &=& \frac{1}{2} f(t).
\eea 

Now, we can rewrite (\ref{modesol}) in a familiar form

\bea\label{quantumlangevin22}
\dot{\hat{c}}_i(t)&=& -i\omega_i \hat{c}_i(t)-i\!\sum\limits_{j} g_{ij} \hat{c}_j(t) -\! \sum_j \frac{\sqrt{\gamma_i\gamma_j}}{2} \hat{c}_j(t) - \sqrt{\gamma_i} \hat{b}_{\rm in}(t)\nonumber \\
\eea where the single decay term in the traditional Langevin equation (\ref{langevinwrong}) has been replaced with a sum over all of the modes $j$ (including the mode $i$ itself). This generalized Langevin equation for multiple bosonic modes is generic, and remains unchanged if the decay rates $\gamma_i$ and frequencies $\omega_i$ are made time-dependent, making the equation amenable to study for experiments involving specific temporal modes of light \cite{Glauber66,Molmer2020,ProppPOVM}, as well as realistic noisy systems in the lab\footnote{Note however, that the correlations between modes in \ref{correlations} do not remain unchanged when $\gamma_i$ and $\omega_i$ are made time-dependent; there are additional convolution terms that disrupt the exact correlations.}.

We note that the first, second, and fourth term in(\ref{quantumlangevin22}) are standard in the quantum Langevin treatment: the first term corresponds to the free evolution of the discrete mode, as its phase rotates with frequency $\omega_i$, the second term describes the interaction between different discrete modes (recall that we define $g_{ii}=0\,\forall i$), and the fourth term contains the input to the network: a mean-zero random variable, once a partial trace is taken over the environment \cite{gardiner00}. It is the third term in the generalized quantum Langevin equation that is of particular interest here; in addition to the standard damping term $-\frac{\gamma_i}{2} \hat{c}_i$ one expects in the quantum Langevin equation, the sum contains additional damping terms proportional to other discrete mode operators. Although these terms relate modes operators to one another like a standard coupling term, they are neither contained in the Hamiltonian, nor do they appear in (\ref{quantumlangevin22}) like a physical coupling $g_{ij}$ (they're missing a $-i$). Indeed, these additional terms are absent in the original work by Gardiner and Collet \cite{input1985}, and might be surprising for the following reason:

Taking $g_{ij}= 0\,\forall i$, the generalized quantum Langevin equation (\ref{quantumlangevin22}) describes an uncoupled network of discrete modes. Although the discrete modes all couple to the same continuum, that continuum is Markovian. As such, one might think that no energy or information can pass between modes. However, the third term in (\ref{quantumlangevin22}) remains in this limit, and encodes quantum correlations between discrete modes (that is, collective decay\footnote{We note that the collective decay here described is distinct from the collective coupling discussed in Refs.~\cite{latune_energetic_2019,latune_heat_2019,latune_thermodynamics_2019,latune_collective_2020}. There, each mode is indistinguishable to the continuum (they have the same energy and decay rates) resulting in a a factorized interaction Hamiltonian and a collective all-to-all coupling. This results in non-trivial finite-temperature thermodynamic effects in addition to non-Markovian evolution. Notably, the collective decay here only induces pair-wise interactions and correlations between modes (\ref{correlations}). Whether this also can lead to interesting quantum thermodynamics remains to be seen.}) that are \emph{inevitable} whenever multiple modes couple to the same continuum. To see this, consider the Fourier transform of (\ref{quantumlangevin22})

\bea\label{quantumlangevinFT}
-i\Delta_i\hat{c}_i(\omega)&=& -\! \sum_j \frac{\sqrt{\gamma_i\gamma_j}}{2} \hat{c}_j(\omega) - \sqrt{\gamma_i} \hat{b}_{\rm in}(\omega)\nonumber \\
\eea where we have let $g_{ij}\rightarrow 0$ and defined a frequency-dependent detuning $\Delta_i = \omega-\omega_i$. Since the sum in the right hand side runs over all $j$ (including the mode $i$), the only $i$-dependence on the right hand side is in the common factor $\sqrt{\gamma_i}$. Dividing both sides by this leaves the right hand side completely $i$-independent, so that we can conclude

\bea\label{correlations}
\frac{\Delta_i \hat{c}_i(\omega)}{\sqrt{\gamma_i}}&=& \frac{\Delta_j \hat{c}_j(\omega)}{\sqrt{\gamma_j}},\,\forall i,j.
\eea 

Taking the expectation value of (\ref{correlations}), we see that the amplitude of a discrete mode is entirely suppressed when the input field is on-resonance with another discrete mode\footnote{Indeed, we see in (\ref{correlations}) that the spectrum of one mode $\hat{c}_i(\omega)$ completely determines every other Heisenberg operator's spectrum $\hat{c}_j(\omega)$ except at the singular frequency, $\omega_j$. Since the same is true for their Hermitian conjugates, we can conclude that the electric field amplitudes $\hat{E}_i\sim \hat{c}_i + \hat{c}_i^\dagger$ of each mode are also maximally correlated. However, this does not hold for higher order moments of the electric field or mode operators; for example, in (\ref{NumberEvolution}) the evolution of the number operator has a (non-trivially) mode-dependent driving term $-\sqrt{\gamma_i}\left({\hat{b}^\dagger_{\rm in}(t) \hat{c}_i(t) + \hat{c}^\dagger_i(t) \hat{b}_{\rm in}(t)}\right)$.}: if $\Delta_i=0$, then $\expect{ \hat{c}_j(\omega)}=0\,\forall j\neq i$. This effect becomes more sharply peaked as the decay rates (and thus spectral widths of the discrete modes) decrease. Again, this is surprising! On-resonance is in general not a sufficient condition for perfect absorption by a qubit (or transmission through a quantum network \cite{ProppNet}), yet it is sufficient for the complete suppression of all other discrete mode amplitudes! Furthermore, when two discrete modes become exactly degenerate, one of them decouples from the continuum entirely as discussed in \cite{Mazzola2009}. This is a result of the unitarity constraint: if two discrete modes are exactly degenerate, they must couple to different continua with different quantum numbers, and will be discussed in more detail in section IV where we study the subsystem dynamics and memory kernel. 


Using the correlations in (\ref{correlations}), we note that we can exactly solve for the behavior of each mode individually in the spectral domain 

\bea\label{langevinspectral}
 -i\Delta_i \hat{c}_i(\omega)&=& -\left(\! \sum_j \frac{\sqrt{\gamma_i\gamma_j}}{2}\frac{\Delta_i}{\Delta_j}\right) \hat{c}_i(\omega) - \sqrt{\gamma_i} \hat{b}_{\rm in}(\omega).\nonumber \\
\eea for the case of no direct coupling ($g_{ij}=0\,\forall ij$) (as well as the special case of $g_{ij} \propto \sqrt{\gamma_i} \,\forall i$\footnote{This corresponds to two modes coupled indirectly through their evanescent fields.}). More generally, if there is a direct coupling between $N$ discrete modes $\hat{c}_i(t)$, one can always rediagonalize the system into a set of $N$ uncoupled modes $\tilde{\hat{c}}_i(t)$, leading again to (\ref{langevinspectral}) but with modified decay rates and resonances. Similarly, one can imagine doing the same thing here: treating the extra decay terms as complex couplings and diagonalizing the system to remove them. This is undesirable for our analysis, firstly because the new modes are no longer eigenmodes of the Hamiltonian (unless they are degenerate in energy), and secondly because our analysis leaves the resonance structure and decay rates of the discrete modes individually manipulable, even allowing for complete decoupling from the environment\footnote{For examples of the diagonalization process, see Refs.~\cite{cardimona1983,altenmuller1995}; there diagonalizing indeed makes more sense as the coupling to the continuum is always present, so that it is not possible to access the ``bare'' atomic states. Here, we have .}.




Returning to the derived generalized quantum Langevin equation (\ref{quantumlangevin}), we note the following: though we have solved for the evolution of the mode amplitude operators $\hat{c}_i(t)$, this is inconvenient for studying the full evolution of the mode number operators themselves $\hat{N}_i(t) = \hat{c}^\dagger_i(t)\hat{c}_i (t)$ \footnote{To get the right open system dynamics for $\hat{N}_i(t)$ at finite temperature from $\hat{c}_i(t)$ alone, one must use the Stratonovich  calculus (which includes the integration endpoints) when evaluating the formal integration so that $\expect{\hat{b}^\dagger_{\rm in}(t) \hat{c}_i(t) + \hat{c}^\dagger_i(t) \hat{b}_{\rm in}(t)}$ in (\ref{NumberEvolution}) contains implicitly the second-order noise term $\gamma_i N_{th}$ \cite{gardiner00}. That calculating $\hat{c}^\dagger_i(t)\hat{c}_i (t)$ from the equations of motion for $\hat{c}_i(t)$ as in (\ref{quantumlangevin}) gives the right zero temperature dynamics for $N_i(t)$ is a convenient result of unitarity of the combined system-environment dynamics; this would not be true of an analysis using the adjoint representation, for instance.}, nor the behavior of their expectation values at finite temperature. This is due to the failure of the quantum Langevin equation to give explicitly the correct noise terms for higher-order operators \footnote{Indeed, it is straightforward to see from Eq. 2.12 of Ref.~\cite{input1985} that the quantum Langevin approach can \emph{only} ever give the lowest order noise terms explicitly; higher order noise terms are contained implicitly and must be extracted by use of the Stratanovich calculus  \cite{gardiner00}.}; in general, one must use the quantum regression theorem to derive the equations of motion that the expectation value of $\hat{N}_i(t)$ follows \cite{Lax1966,Swain1981,gardiner00,steckquoptics,Blocher2019}. However, here we use a different trick\footnote{Note that the trick we use in (\ref{infinitesimal}) is in the style of the Ito calculus  \cite{gardiner00}, so that we can later separately evaluate system and environment expectation values as usual $\expect{\hat{b}^\dagger_{\rm in}(t) \hat{c}_i(t)} = \expect{\hat{b}^\dagger_{\rm in}(t)}\expect{ \hat{c}_i(t)}$.} to get the correct finite-temperature noise term in the equations of motion for $N_i(t)$:

We can rewrite the generalized quantum Langevin equation (\ref{quantumlangevin22}) as an infinitesimal expansion in dt

\begin{widetext}\bea\label{infinitesimal}
\hat{c}_i(t+dt) &=& \hat{c}_i(t) + dt\left(
-i\omega_i \hat{c}_i(t)-i\!\sum\limits_{j} g_{ij} \hat{c}_j(t) -\! \sum_j \frac{\sqrt{\gamma_i\gamma_j}}{2} \hat{c}_j(t) - \sqrt{\gamma_i} \hat{b}_{\rm in}(t) \right)
\eea\end{widetext} and expand the infinitesimally evolved number operator

\begin{widetext}\bea\label{infinitesimal2}
\hat{N}_i(t+dt) \equiv  \hat{c}_i^\dagger(t+dt) \hat{c}_i(t+dt) &=&  \hat{c}_i^\dagger(t)\hat{c}_i(t) +
-i\, dt\!\sum\limits_{j\neq i} g_{ij} \left(\hat{c}_i^\dagger(t)\hat{c}_j(t)-\hat{c}_j^\dagger(t)\hat{c}_i(t)\right)  \\
&-&\! dt \sum_j \frac{\sqrt{\gamma_i\gamma_j}}{2} \left(\hat{c}_i^\dagger(t)\hat{c}_j(t)+\hat{c}_j^\dagger(t)\hat{c}_i(t)\right)  - dt\sqrt{\gamma_i} \left(\hat{b}^\dagger_{\rm in}(t)\hat{c}_i(t)+\hat{c}_i^\dagger(t)\hat{b}_{\rm in}(t)\right) \nonumber \\
&+& dt^2 \gamma_i \hat{b}_{\rm in}^\dagger (t) \hat{b}_{\rm in} (t) + \mathcal{O}(dt^2)\nonumber
\eea\end{widetext} where the reader will notice we have kept a single term of order $dt^2$. While this term is second order in the evolution of the quantum operator, it reduces to first order in calculation of the expectation value as, under the Markov approximation, $dt\expect{\hat{b}_{\rm in}^\dagger (t) \hat{b}_{\rm in}(t)}=N_{\rm th}$, with $N_{\rm th}$ the average number of  excitations at the resonance energy which are incident upon the system from the continuum. From this expression, we can derive Heisenberg equations of motion for the expectation value\footnote{The associated finite-temperature Heisenberg equations of motion for the number operators themselves (as opposed to their expectation values) are intrinsically stochastic and cannot be calculated deterministically.} of the number operator 

\begin{widetext}\bea\label{NumberEvolution}
\!\!\!\!\!\!\!\!\!\!\!\!\expect{{\dot{\hat{N}}_i(t)}}&=& -\gamma_i \expect{\hat{N}_i(t)} - i\sum_{j\neq i}g_{ij}  \left(\expect{\hat{c}_i^\dagger(t) \hat{c}_j(t)  - \hat{c}_j^\dagger(t) \hat{c}_i(t)}\right) - \sqrt{\gamma_i} \sum_{j\neq i} \frac{\sqrt{\gamma_j}}{2} \left(\expect{\hat{c}_i^\dagger(t) \hat{c}_j(t)  + \hat{c}_j^\dagger(t) \hat{c}_i(t)}\right) \nonumber\\
&-& \sqrt{\gamma_i}\left(\expect{\hat{b}^\dagger_{\rm in}(t) \hat{c}_i(t) + \hat{c}^\dagger_i(t) \hat{b}_{\rm in}(t)}\right) + \gamma_i N_{th}  \nonumber \\
&\equiv&  -\gamma_i \expect{\hat{N}_i(t)} - \sum_{j\neq i}g_{ij}  \expect{\hat{Y}_{ij}(t)} - \sum_{j\neq i} \frac{\sqrt{\gamma_i\gamma_j}}{2}\expect{\hat{O}_{ij}(t)} \nonumber - \sqrt{\gamma_i}\left(\expect{\hat{b}^\dagger_{\rm in}(t) \hat{c}_i(t) + \hat{c}^\dagger_i(t) \hat{b}_{\rm in}(t)}\right) +\gamma_i N_{th} \nonumber \\.
\eea\end{widetext}

In the first line of (\ref{NumberEvolution}), the first term corresponds to decay, the second term correspond to inter-mode coupling (which we are not interested in and will set to zero shortly), the third term corresponds to coherent driving from the continuum (or stochastically, a mean-zero random variable that will disappear upon taking an expectation value), the fourth term describes the correlations the modes develop through their connection to a shared continuum, and the last term is incoherent thermal driving as discussed in the paragraph above. In the second line we define two new Hermitian operators: $\hat{O}_{ij}(t) \equiv \hat{c}_i^\dagger(t) \hat{c}_j(t)  + \hat{c}_j^\dagger(t) \hat{c}_i(t)$ and $\hat{Y}_{ij}(t) \equiv i( \hat{c}_i^\dagger(t) \hat{c}_j(t)  - \hat{c}_j^\dagger(t) \hat{c}_i(t))$. These obey their own Heisenberg equations of motion with expectation values 

\begin{widetext}\bea\label{auxEqs|:1}
\!\!\!\!\!\!\!\!\!\!\!\!\!\!\!\!\!\!\!\!\!\!\!\!\!\!\!\expect{\dot{\hat{O}}_{ij}(t)} =  (\omega_i-\omega_j) \expect{\hat{Y}_{ij}(t)} &-& \sum_k \left(g_{ik}\expect{\hat{Y}_{jk}(t)} + g_{jk}\expect{\hat{Y}_{ik}(t)}\right) - \sum_k \left(\frac{\sqrt{\gamma_i \gamma_k}}{2} \expect{\hat{O}_{jk}(t)}  + \frac{\sqrt{\gamma_j \gamma_k}}{2} \expect{\hat{O}_{ik}(t)} \right) \\
&-&\sqrt{\gamma_i}\left(\expect{\hat{b}^\dagger_{\rm in}(t) \hat{c}_j(t) + \hat{c}^\dagger_j(t) \hat{b}_{\rm in}(t)} \right)-\sqrt{\gamma_j}\left(\expect{\hat{b}^\dagger_{\rm in}(t) \hat{c}_i(t) + \hat{c}^\dagger_i(t) \hat{b}_{\rm in}(t)} \right) +  2\sqrt{\gamma_i\gamma_j} N_{th}\nonumber\\
\nonumber \\ 
\!\!\!\!\!\!\!\!\!\!\!\!\!\!\!\!\!\!\!\!\!\!\!\!\!\!\!\expect{\dot{\hat{Y}}_{ij}(t)} =  (\omega_i-\omega_j) \expect{\hat{O}_{ij}(t)} &+& \sum_k \left(g_{ik}\expect{\hat{O}_{jk}(t)} - g_{jk}\expect{\hat{O}_{ik}(t)}\right) +\sum_k \left(\frac{\sqrt{\gamma_i \gamma_k}}{2} \expect{\hat{Y}_{jk}(t)}  - \frac{\sqrt{\gamma_j \gamma_k}}{2} \expect{\hat{Y}_{ik}(t)} \right) \nonumber \\ \label{auxEqs|:2}
&-&i\sqrt{\gamma_i}\left({\hat{b}^\dagger_{\rm in}(t) \hat{c}_j(t) - \hat{c}^\dagger_j(t) \hat{b}_{\rm in}(t)} \right)+i\sqrt{\gamma_j}\left({\hat{b}^\dagger_{\rm in}(t) \hat{c}_i(t) - \hat{c}^\dagger_i(t) \hat{b}_{\rm in}(t)} \right).\nonumber \\
\eea\end{widetext}

The second auxiliary equation (\ref{auxEqs|:2}) has the form of a discrete site particle number current density\footnote{This is clear from how we have defined $\hat{Y}_{ij} \equiv i( \hat{c}_i^\dagger \hat{c}_j  - \hat{c}_j^\dagger \hat{c}_i)$ in terms of a difference of two terms, the first moving an excitation from the $i$th mode to the $j$th mode and the second doing the inverse.}, where the coherences between discrete modes (characterized by the first auxiliary equation (\ref{auxEqs|:1}) for $\hat{O}_{ij}(t)$) drive energy exchange between modes. We pause to note that collective decay is a unique feature of open quantum systems; closed systems require direct couplings to enable energy and information exchange, whereas open systems can accomplish this through their decay processes.

In the absence of a coherent input field $\left(\expect{\hat{b}_{\rm in}(t)}=0\right)$ and at zero temperature ($N_{\rm th} = 0$), the set of equations (\ref{NumberEvolution}--\ref{auxEqs|:2}) describing the evolution of the number operator and the two auxiliary operators form a closed system of equations, which can be written as a Gramm matrix. This can be solved using the standard methods of matrix eigenvalue decomposition, yielding (in the simplest case of two discrete modes) an exact solution determined from initial conditions at $t=0$. Note that in the absence of initial coherences $\expect{\hat{O}_{ij}(0)}=\expect{\hat{Y}_{ij}(0)}=0$. For the case study of two non-interacting bosonic modes, we solve the equations of motion for the number operator, auxiliary operators, and discrete mode amplitude operators numerically using the computational package \texttt{Mathematica}, as detailed in Appendix B.

\section{Subsystem Dynamics}

   \begin{figure}[h] 
   \centering
	\includegraphics[width=\linewidth]{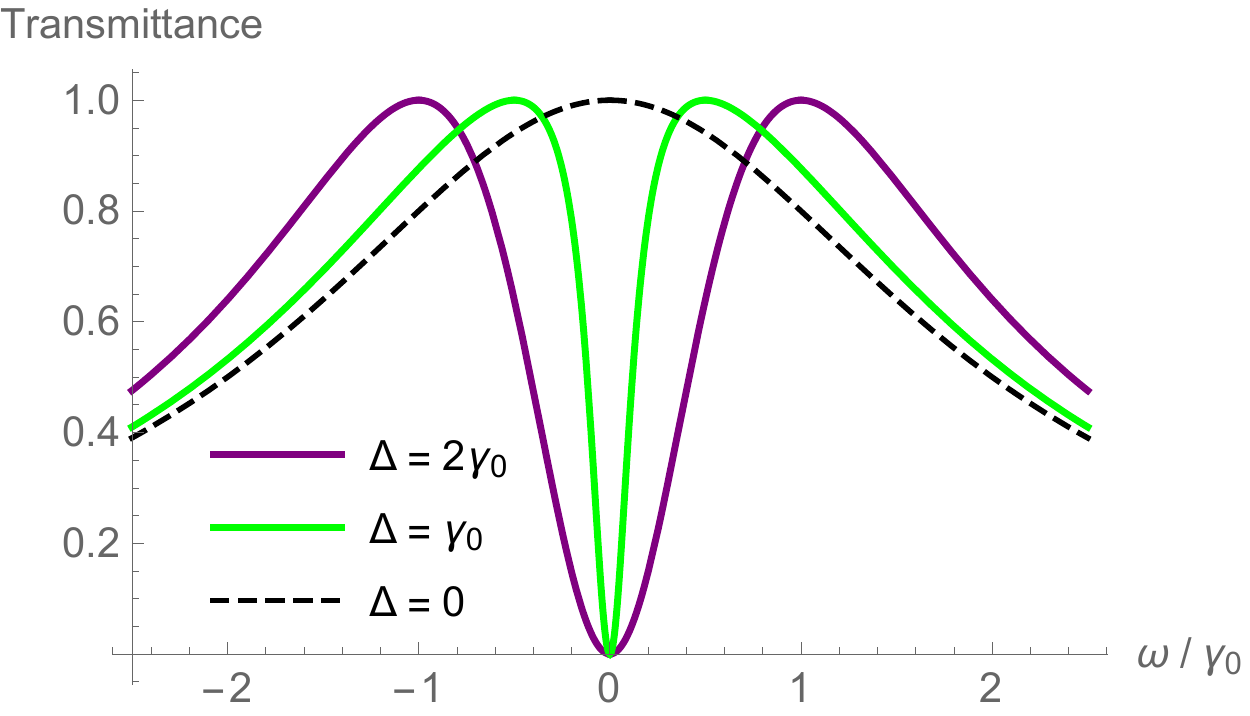} 
	\caption[]{Frequency-dependent transmission through a two-sided network comprised of two discrete modes detuned by $\Delta$, with uniform decays from both modes to both continua ($\gamma_1=\gamma_2$) and frequency measured with respect to the average $\frac{\omega_1+\omega_2}{2}$.}
	\label{DetuningPlotlot}
\end{figure}

Before moving on to our numerical study of the two-mode system, we perform one more calculation to better understand the dynamics of the single-mode subsystem.

We can make use of the strong correlations between modes to integrate out one mode analogously to as we did for the continuum in our derivation of the quantum Langevin equation in (\ref{quantumlangevin22}). In the absence of direct coupling ($g_{ij}=0$), we perform this operation for an $N$-mode parallel network as follows:

In the quantum Langevin equation in the absence of coupling, the time evolution of a mode $\hat{c}_i(t)$ depends on the other $N-1$ modes (and itself) via a term $-\sum_j \frac{\sqrt{\gamma_i\gamma_j}}{2} \hat{c}_j(t)$. However, in the frequency domain, we can make use of the strong correlations between the discrete mode amplitude operators in (\ref{correlations}), resulting in the spectral solution for the single modes alone (\ref{langevinspectral}). Using the method of residues, we can move back to the time domain and arrive at a time-convolution-full (as opposed to time-convolutionless) Langevin equation for a single mode

 \bea\label{quantumlangevin2}
\!\dot{\hat{c}}_i(t)&=& -i\omega_i \hat{c}_i(t)- (\tilde{\gamma}_i * \hat{c}_i)(t) - \sqrt{\gamma_i} \hat{b}_{\rm in}(t)
\eea  where we have defined a time-dependent memory kernel (the generalization of a time-dependent decay rate)

\bea\label{memorykernal}
\!\!\!\!\! \!\!\!\  \tilde{\gamma}_i(t)\equiv\frac{\gamma_i}{2} 2\pi \delta(t) +\sum\limits_{j\neq i} \frac{\gamma_j}{2} \left(2\pi\delta(t) + i\pi(\omega_j-\omega_i)e^{i\omega_j t} {\rm Sign}[t]\right). \nonumber \\ 
  \eea

We note that the memory kernel (\ref{memorykernal}) is highly time non-local, except for when the resonance energies become degenerate. In this case, all that remains is a modification of the decay rate. This effect can be observed in spectral transmittance through a two-sided network comprised of two uncoupled modes in parallel, as studied in Ref.~\cite{ProppNet}. When the two modes become exactly degenerate, a frequency of maximum transmission and a frequency of perfect reflection annihilate, leaving one state with a modified decay rate (and thus spectral width) and the other discrete mode perfectly decoupled from the continuum as shown in Fig. \ref{DetuningPlotlot}. Indeed, this is necessary to preserve unitary, and is precisely the dark state generation observed and discussed in Ref.~\cite{Mazzola2009}.

 \begin{figure}[b] 
	\includegraphics[width=1\linewidth]{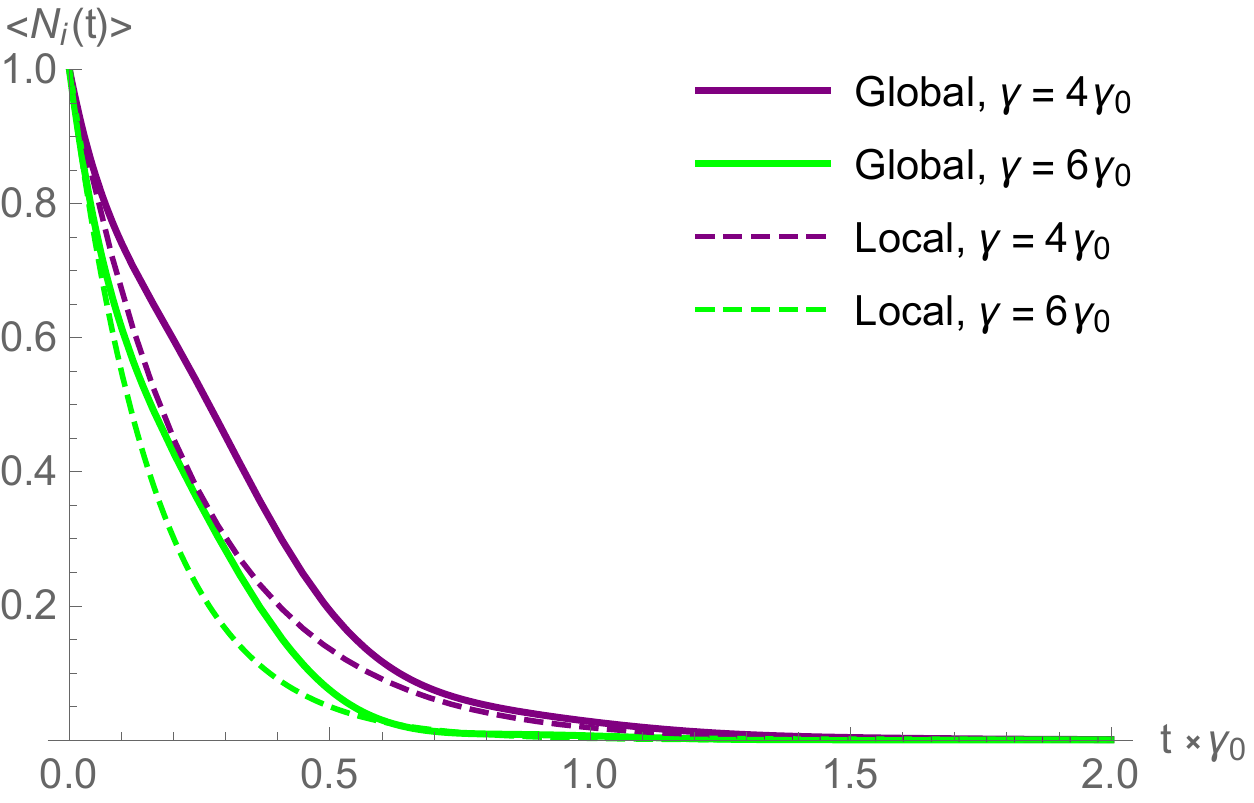} 
	\caption[]{Time-dependent number operator expectation value (excited state population) for two modes far from degeneracy ($\omega_1-\omega_2\equiv\Delta = 10\gamma_0$), both starting in the excited state. Here and elsewhere, the slower decaying state (purple) is above the faster decaying state (green).}
	\label{FockStatesFar}
\end{figure}

From the highly time-non-local nature of the memory kernel in (\ref{memorykernal}), it is tempting to conclude from this alone the general behavior is non-Markovian; after all, a time-non-local memory kernel for the evolution of quantum states is a hallmark of non-Markovianity \cite{Kossakowski2010}. However, this memory kernel is in a Heisenberg equation of motion for a system operator, and not a master equation for density matrix evolution. Modern definitions of non-Markovian dynamics for open quantum systems rely on properties of density matrices (i.e. a monotonically decreasing trace distance, for a thorough review see  Ref.~\cite{breuer2016} and are geared towards \sch or interaction picture analyses (although there is already some work on Heisenberg picture analyses, see Ref.~\cite{Vega2017,Li2018}). The task of formally connecting the existing state-centric measures of non-Markovianity to state-independent Heisenberg picture studies is left for future work.

To observe memory effects (including energy back flow \cite{Piilo2009,Vacchini2016}) we must look at a specific example, which is the task we turn to now.

 \begin{figure}[t] 
	\includegraphics[width=1\linewidth]{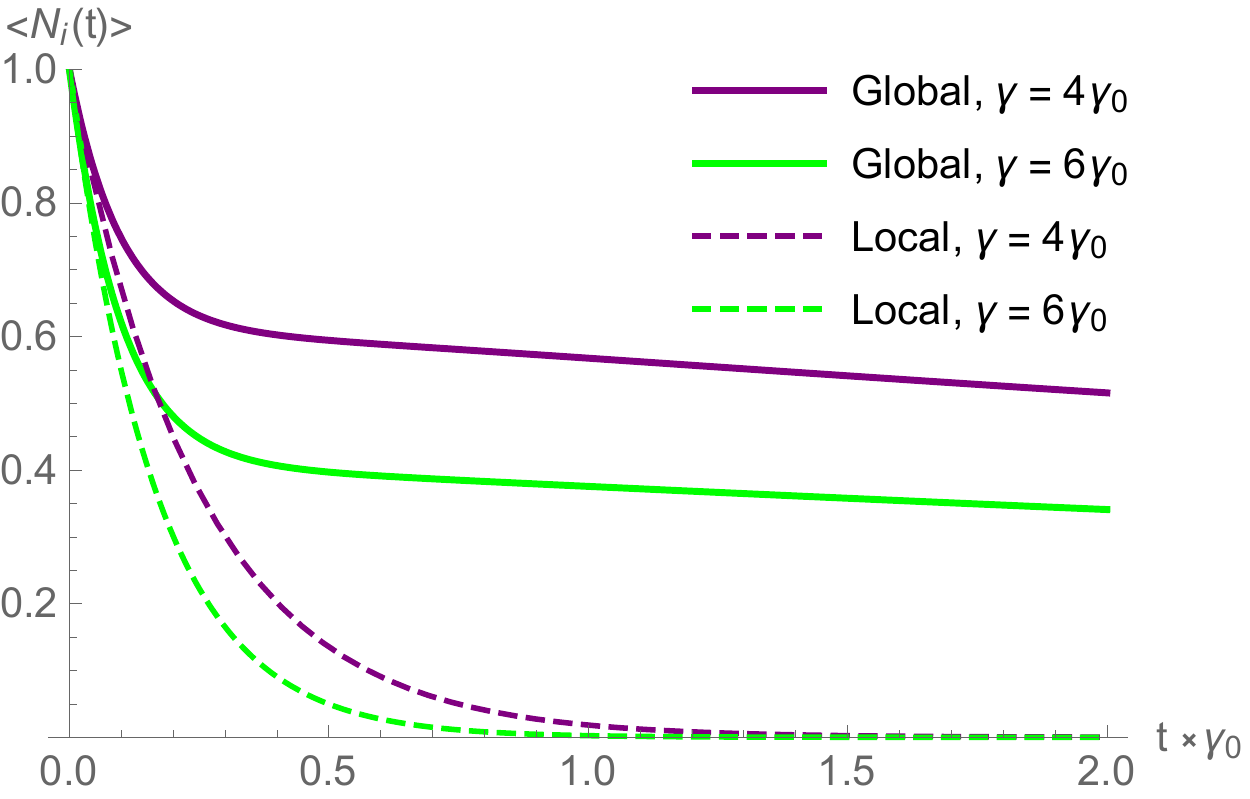} 
	\caption[]{Time-dependent number operator expectation value (excited state population) for two near-degenerate modes ($\Delta=1\gamma_0$), both starting in the excited state.}
	\label{FockStatesClose}
\end{figure}

 \begin{figure}[b] 
	\includegraphics[width=1\linewidth]{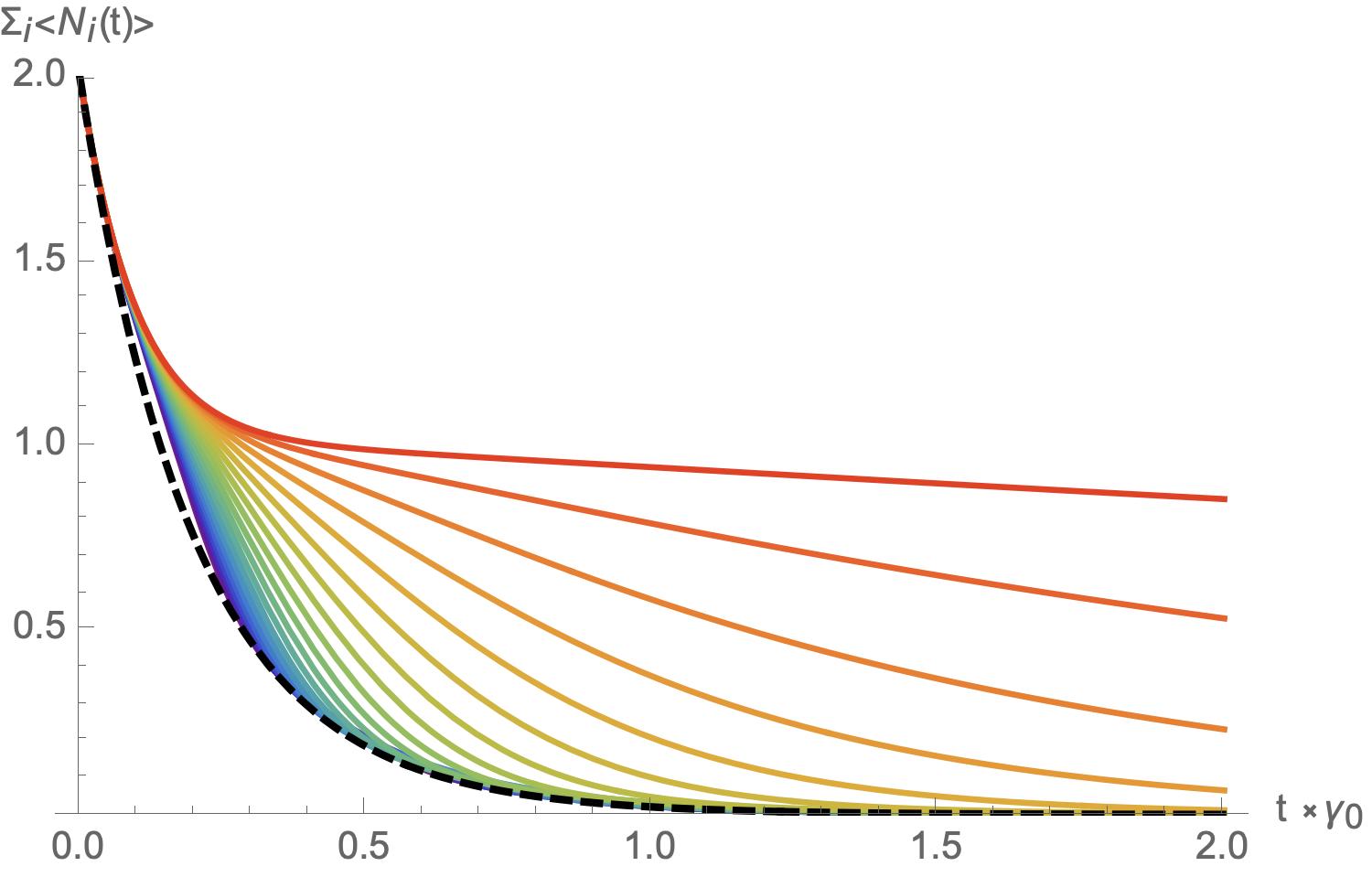} 
	\caption[]{Total time-dependent number operator expectation value (total excited state population) for two modes decaying at rates $\gamma_1=4\gamma_0$ and $\gamma_2=6\gamma_0$ to a shared global continuum, plotted as a function of time for a variety of detunings $\Delta$. These range from $\Delta=1\gamma_0$ (top, solid, red) to $\Delta=20\gamma_0$ (bottom, solid, violet). Also plotted is the total excited state population for two independently (locally) decaying modes (black, dashed), corresponding to the infinite detuning limit $\Delta\rightarrow \infty$.}
	\label{FockTotalNumber}
\end{figure}

\section{Example: Two modes}

We will now focus on the case we are most interested in: two uncoupled ($g_{ij}=0$) discrete modes (e.g. cavity modes, qubits in the weak-field limit) decaying to a continuum of states with no initial excitations or coherences in the external field $\left(\expect{\hat{b}_{\rm in}(t)}=0\right)$ under the first Markov approximation $\left(\frac{d\gamma_i}{d\omega} = 0\,\forall i\right)$.

 \begin{figure}[t] 
	\includegraphics[width=1\linewidth]{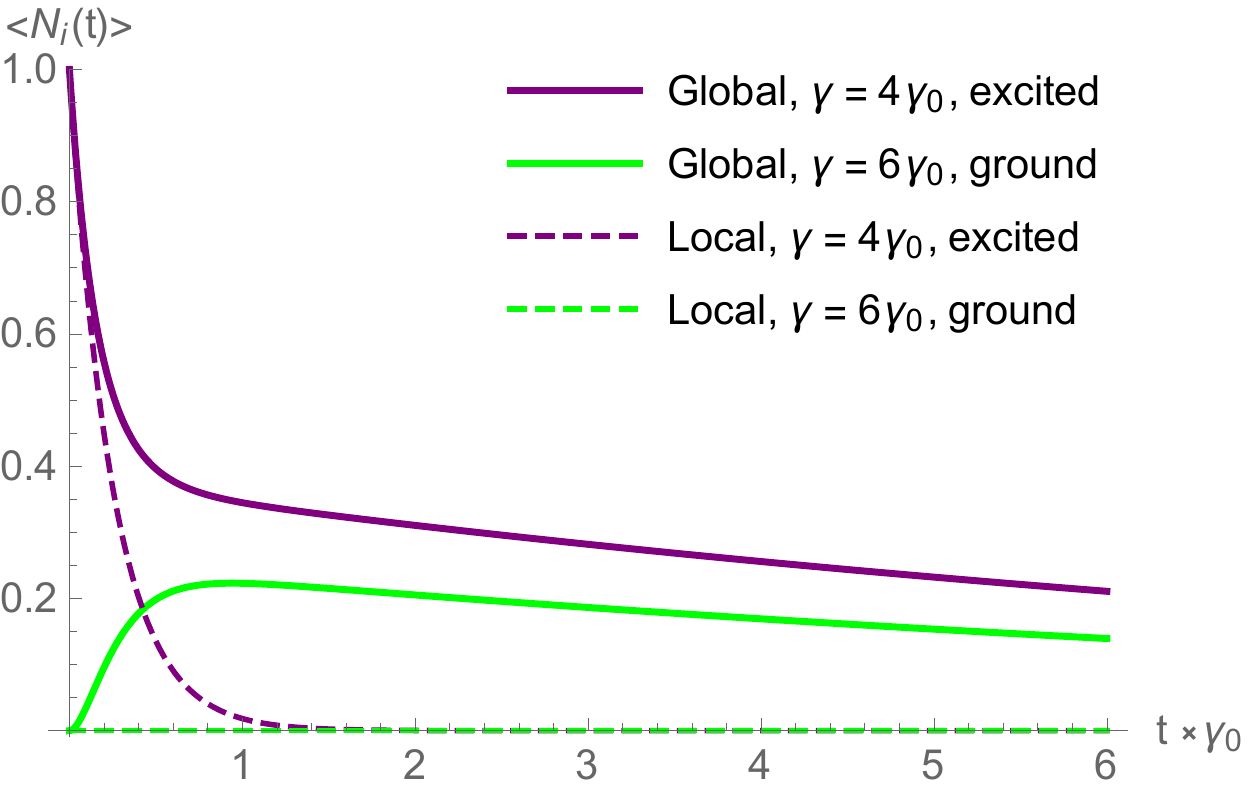} 
	\caption[]{Time-dependent number operator expectation value (excited state population) for two near-degenerate modes ($\Delta = 1\gamma_0$), with the one mode starting in the ground state and one mode starting in the Fock state with a single excitation (the first excited states).}
	\label{FockGroundExcited}
\end{figure}

 \begin{figure}[b] 
	\includegraphics[width=1\linewidth]{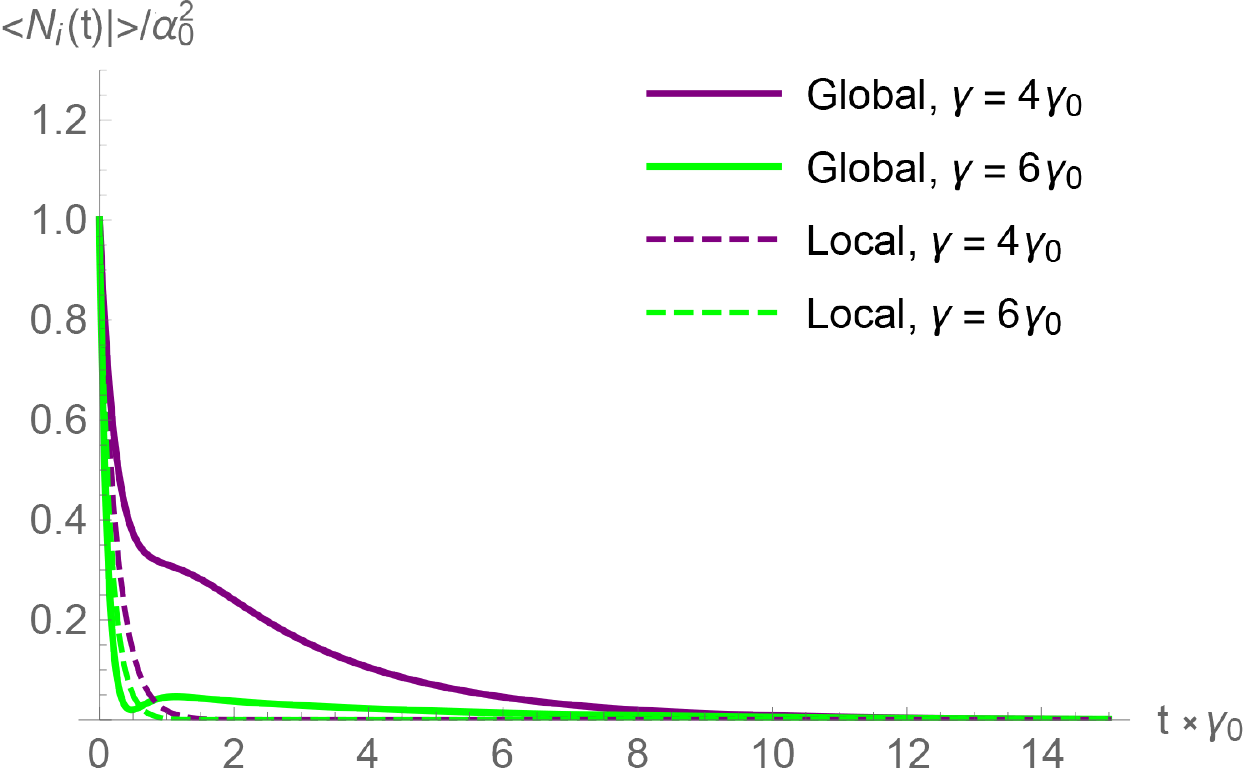} 
	\caption[]{Time-dependent discrete mode population for two bosonic modes, both starting in the same coherent state $\ket{\alpha_0}$ with the same phase and mode detuning $\Delta = 2\gamma_0$. Note that the behaviors of the discrete mode populations are invariant under rescaling by $\alpha_0^2$.}
	\label{BosonicNumberSame}
\end{figure}

First, we study two modes each initialized to the Fock state with a single excitation (for fermions, the excited state). When the two modes are distantly spaced in energy as in Fig. \ref{FockStatesFar}, we observe only a slight sub-radiant deviation between the cases where the modes decay globally to a shared continuum, or locally to their own continua. As the two modes approach each other in energy, this subradiance becomes more dramatic and the lifetimes of both modes are extended substantially (Fig. \ref{FockStatesClose}). Similarly, the effect on the total occupancy of both modes is subradiant, resulting in non-exponential decay of the full system as the energy spacing decreases (Fig. \ref{FockTotalNumber}). When the degeneracy becomes exact, one mode decouples from the continuum completely as made clear in (\ref{memorykernal}) and observed experimentally as a dark state in \cite{pellegrino2014}.

 \begin{figure}[b] 
	\includegraphics[width=1\linewidth]{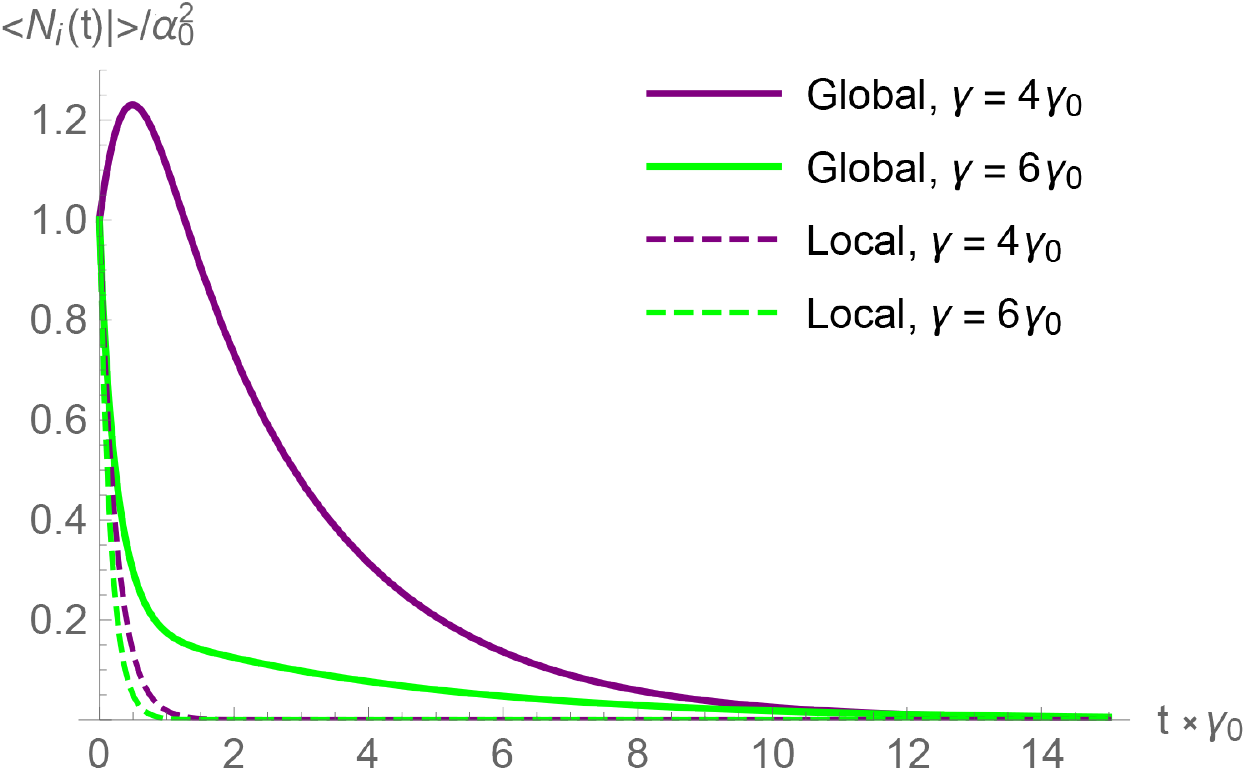} 
	\caption[]{Time-dependent discrete mode population for two bosonic modes, both starting in the a coherent state with same magnitude $\alpha_0$ but with opposite phases and mode detuning $\Delta = 2\gamma_0$.}
	\label{BosonicNumberDifferent}
\end{figure}

We also consider the case where one mode is initialized to a Fock state with a single excitation, and the other mode is initialized in the ground state (Fig. \ref{FockGroundExcited}). In this case, we observe the same subradiance for the qubit that starts in the excited state. However, the qubit starting in the ground state gains a substantial population before decaying. This is evidence of genuine energy transfer between the two modes, despite both there being no direct coupling between them and their individual decays to the shared global continuum being Markovian.

In the three cases studied so far, the expectation value of the number operator never exceeds unity for either mode. Hence, the numerical results plotted previously will apply to fermionic qubits as well as bosonic modes. Now, we will restrict our study to bosons and study what happens when we initialize each mode in a coherent state.

 \begin{figure}[t] 
	\includegraphics[width=1\linewidth]{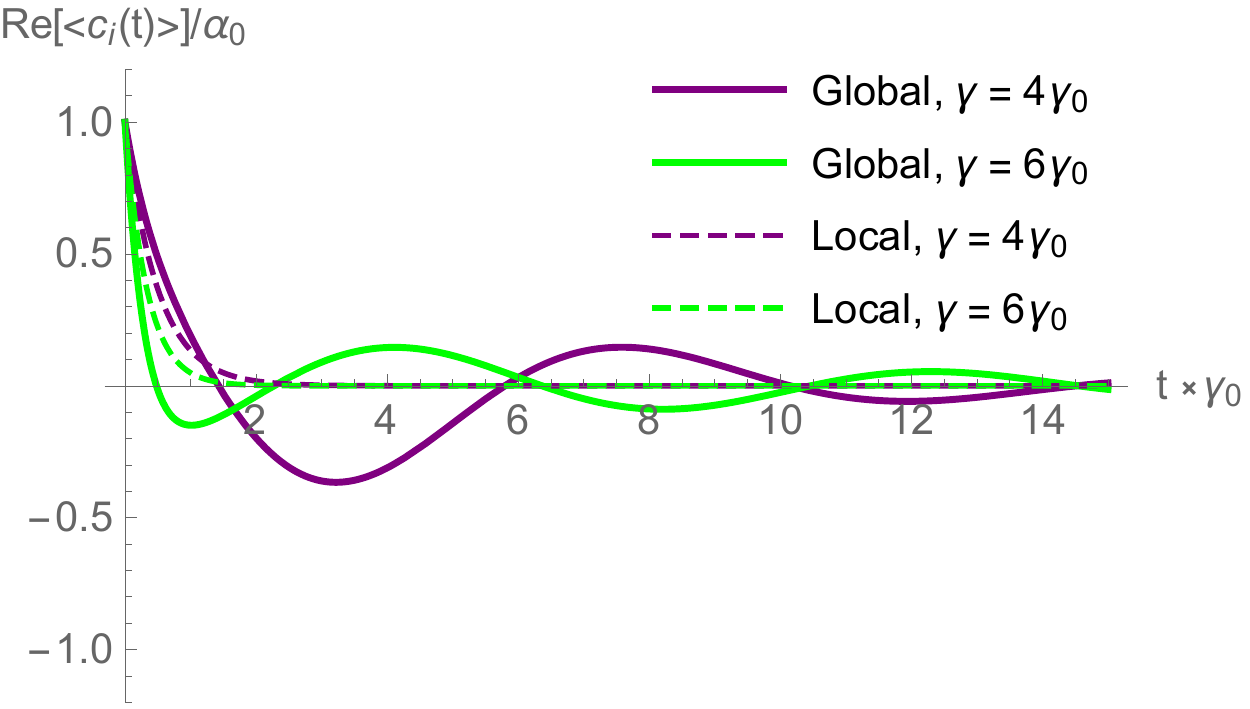} 
	\caption[]{Time-dependent electric field amplitude for two bosonic modes, both starting in the same coherent state $\ket{\alpha_0}$ with the same phase and mode detuning $\Delta = 2\gamma_0$. Note that the behaviors of the electric field amplitudes are invariant under rescaling by $\alpha_0$.}
	\label{BosonicElectricSame}
\end{figure}

 \begin{figure}[b] 
	\includegraphics[width=1\linewidth]{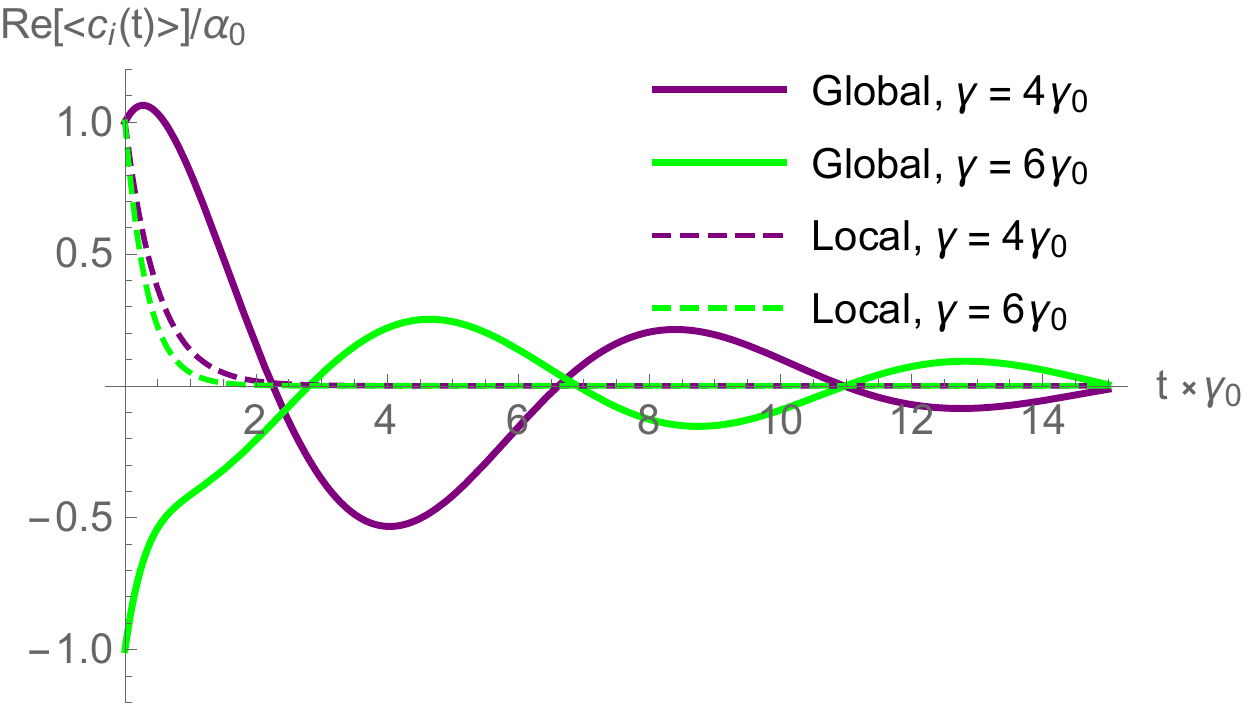} 
	\caption[]{Time-dependent electric field amplitude for two bosonic modes, both starting in the a coherent state with same magnitude $\alpha_0$ but with opposite phases and mode detuning $\Delta = 2\gamma_0$.}
	\label{BosonicElectricDifferent}
\end{figure}

For an initial coherent state, there is an additional operator of interest: the electric field amplitude. While a measurement of the single-mode electric field operator  $\hat{E}_i\sim \hat{c}_i + \hat{c}_i^\dagger$ is always zero for Fock states (and thus not of interest), here it is non-zero and exhibits phase-sensitivity: similarly to the photon counting measurements in Figs. \ref{BosonicNumberSame} and \ref{BosonicNumberDifferent}, for two in-phase initial coherent states we observe that at short times one mode is sub-radiant while the other is super-radiant with both exhibiting subradiance at longer times (Fig. \ref{BosonicElectricSame}), and for two out-of-phase initial coherent states we observe the faster decaying mode drives the slower decaying modes electric field amplitude (Fig. \ref{BosonicElectricDifferent}).

First, we notice a change between initial super-and sub-radiant decay depending on the relative phase between the two initial coherent states: when the modes start in phase with each as in Fig. \ref{BosonicNumberSame}, we observe initial subradiance for one mode and superradiance for the other, eventually resulting in population revival in the faster decaying state and subradiance for both modes at longer times; when the modes start out of phase with each other as in Fig. \ref{BosonicNumberDifferent}, the result is subradiance for both modes as in the Fock state case, but with the faster decaying mode driving the population of the other mode. It is the case where both states are sub-radiant that leads to a population increase in the slower decaying mode. This is easily explained as a result of wave-interference; since the coherent states are initially out of phase, they initially interfere destructively in the continuum and population from the faster decaying mode end up in the slower decaying mode despite their being a) no direct coupling and b) only an mutual driving of the same continuum field---a driving that would be irreversible and incoherent for a single mode alone. Similarly for the in-phase case, initial constructive interference in the continuum results in the faster decaying state exhibiting superradiance (Fig. \ref{BosonicElectricSame}).

Figs. \ref{BosonicNumberSame}–\ref{BosonicNumberDifferent} can also be understood in terms of particles and non-Markovianity; there is a non-zero probability for an excitation leaking out of a  mode to scatter off of an effective structured continuum and return to its original source. This gives rise to energy back-flow, from which we can conclude that the system is evolving in a genuinely non-Markovian fashion \cite{Piilo2009,Vacchini2016}.

\section{Two Applications}

1) Reservoir engineering is a well-known method for modifying the decay and decoherence of a system by ``engineering'' a desired system-environment interaction \cite{poyatos1996}. By modifying enviornmental parameters in addition to (or instead of) system parameters, one can induce a plethora of system behaviors such as noise squeezing \cite{Rabl2004} (and relatedly, amplification  \cite{metelmann2014}), non-reciprocal photo-transmission \cite{metelmann2015}, and generation of entanglement between two subsystems \cite{krauter2011,muschik2012}. A more pedestrian but nonetheless very important application is the stabilization of quantum states \cite{Sarlette2011}. As we have discussed, adding a second auxiliary mode of similar energy will extend the lifetime of a first mode coupled to the same continuum, even if that auxiliary mode is not initially populated (Fig. \ref{FockGroundExcited}). This is a rudimentary and simple to implement form of reservoir engineering; the presence of a second auxiliary mode will act as an environmental memory to the first, slowing down the dissipation of information and energy into a continuum (or photon loss channel \footnote{For example, it is natural to describe a photo detector as a quantum network of discrete states interacting with both a target input continuum and one or more loss channels, also modeled as continua. Photon loss contributes to both inefficiency \cite{ProppNet} and POVM impurity \cite{ProppPOVM}, and can be mitigated with auxiliary mode reservoir engineering.}).

2) Conversely, another application of this phenomenon is purely theoretical. The pseudomode representation is a well-studied and useful calculational tool for understanding the dynamics of non-Markovian open quantum systems, especially in the strong-coupling regime \cite{pseudomodes2,pseudomodes3,pseudomodes4,pseudomodes1,pleasance2021}. In essence, non-Markovianities in the system-environment are absorbed into coupled, potentially fictitious auxiliary system modes through Fano diagonalization. In addition to coupling to the quantum system in question, each pseudomode couples to a number of local Markovian continua. For the simplest example, consider the case of an atom interacting with a continuum through a simple Fabry-Perot frequency filter. Here, the effect in the spectral domain is a Lorentzian suppression of decay to the continuum \cite{siegman86}. This filter-induced non-Markovianity is naturally described by an intermediary mode coupling to the atom directly, and \emph{then} decaying to the flat Markovian continuum (free space). 

More generally, the standard pseudomode method relies on a decomposition of the Fourier transformed decay rate or memory kernel into a sum of approximately Lorentzian and Lorentzian-like distributions. The closer to a Lorentzian profile, the more exact the approximation of the environment as an intermediating pseudomode network becomes. As currently formulated, the Fano-diagonalization process that extracts the pseudomodes from the system-environment coupling is only capable of extracting pseudomodes that couple to each other, to the quantum system of interest, and to their own local continua. Allowing pseudomodes to couple to shared global continua as well could more accurately describe environments with spectra from from Lorentzianity. In the absence of an analytic Fano-diagonalization-like process to extract pseudomodes coupled to shared global continua, machine learning may provide a natural method for decomposing open quantum systems into traditional pseudomode networks augmented by collectively decaying pseudomodes \footnote{It bears mentioning that there are intersections between the two applications proposed here; non-Markovianity can be a resource for reservoir engineering as explored in Refs.~\cite{poggi2017,mirkin2019}, and it is natural to consider the inverse machine learning application: tailoring non-Markovian systems to perform specific reservoir engineering tasks.}. 

\section{Conclusions}

Having studied the mathematical of this phenomenon and explored possible applications, we briefly return our attention to the question of interpretation. At first glance, it is somewhat puzzling that we have applied the first Markov approximation whenever possible in the derivation in the theory section, and yet ended up with population revival and memory effects. To this three remarks are warranted: firstly, population revival within a subsystem does not violate Markovianity for the larger system, and the decay of the total population---while non-exponential---is nonetheless monotonic and Markovian (Fig. \ref{FockTotalNumber}). Secondly, the origin of this effect is the last term of (\ref{modesol}): in pulling both $\kappa_i(\omega)$ and $\kappa_j(\omega)$ outside the integral, we invoke not only the Markov approximation for each mode individually but also for each \emph{pair} of modes. In this way, making the Markov approximation for each state individually conspires to hide a broader assumption giving rise to this collective decay phenomenon, memory effects for the single-mode subsystems, and non-exponential subradiance for the total system. Thirdly, we would like to reiterate that collective decay is well-known to generate non-Markovianity in a number of systems \cite{Fleming2012,breuer2016,Vega2017,Solano2020}. Furthermore, it is known that multi-mode optical cavities can spontaneously generate coherences \cite{Houck2015} and, for an atom coupled to a multi-mode optical cavity, the subsystem dynamics are non-Markovian \cite{Valetin2022}.

This phenomenon also can be understood as a breakdown of both the immediate irreversibility usually associated with Markovian decay for each mode, and the separability of each mode's decay into separate processes in $1$D \footnote{The question of how closely atoms must be spaced to exhibit collective decay was discussed extensively in a recent publication along a similar line to our own (though there the focus is systems with many-body interactions in free space), see Ref.~\cite{masson2021},.}; the correlations between discrete mode amplitudes prevent the collective decay process from being decomposable into separate decays. Since the modes interact with the same quantum field (the electromagnetic continuum of states) at the same (approximate) physical location, there is simply no time-scale fast enough for energy to dissipate from one mode without reaching the other modes. Thus, the second state acts as environmental memory to the first. To consider a classical analogy, consider the two charged oscillators: classically the oscillation of one will drive the electromagnetic field at that position, which inevitably drives the other as well. For the quantum case, this means \emph{there is simply no regime where the Markov approximation can be justified or interpreted in the usual manner}. Invoking it mathematically results in the memory effects observed here. That the induced correlated decay disappears far from degeneracy is consistent with this interpretation (the light emitted from one mode is unlikely to drive another far from resonance), as is the observation that this effect disappears for exactly degenerate states (in $1$D, there simply are no other degrees of freedom and so one mode must decouple to preserve unitarity). 

In conclusion, we have derived and discussed the features of the generalized quantum Langevin equation (\ref{quantumlangevin}), amenable to study of systems with multiple modes decaying to the same continuum. For initially uncoupled modes, spontaneous radiative coupling induces pair-wise correlations between discrete mode amplitudes. For the case study of a two-mode system, this dramatically extends the lifetimes of both modes creating non-exponential total population decay, and also induces population revival within the single-mode subsystems. The lifetime extension makes this phenomenon a natural candidate for reservoir engineering, and the simple non-Markovian structure induced suggest natural incorporation of this phenomenon into previously developed methods for modeling non-Markovian open quantum systems. 

Collective decay phenomena are ubiquitous in quantum optics; by explicating their study within the input-output formalism, we hope to support their future study.


\section{Acknowledgements} 

This work would not have been possible without a number of individuals. This paper is a direct result of a series of productive conversations with Graeme Pleasance; I am very grateful for our chance meeting at the cancelled 2020 APS March Meeting, as well as his expertise on open quantum systems and pseudomodes. I also gratefully acknowledge Steven van Enk, both for his useful comments on this project as well as for first teaching me how to modify input-output theory to incorporate the correlations between modes coupled to the same continuum. I would also like to thank Pablo Solano, Carlton C. Caves, Justin Dressel, Francisco Elohim Becerra, Manuel Munoz, and Philip Blocher for their valuable comments.


This work was supported by National Science Foundation Grant No. PHY-1630114, and is based upon work partially supported by the U.S. Department of Energy, Office of Science, National Quantum Information Science Research Centers, Quantum Systems Accelerator.

\bibliography{virtualcoupling3.bib}

\appendix{
\section{Fermionic Modes}

We briefly sketch the challenge in studying fermionic discrete modes coupled to the same Markovian continuum. We begin by rewriting (\ref{Hmot}) explicitly in terms of time-dependent fermionic raising and lowering operators for the modes $\hat{\sigma}^+_i(t)$ and $\hat{\sigma}^-_i(t)$ instead of bosonic creation and annihilation operators. We write

\bea\label{HFerm}
\!\!\!\!\!\dot{\hat{b}}(t,\omega) &=& -i\omega \hat{b}(t,\omega) + \sum_j \kappa_j(\omega) \hat{\sigma}^-_j (t)\\
\!\!\!\!\!\!\dot{\hat{\sigma}}^-_i(t)&=& -i[\hat{\sigma}^-_i(t),\hat{H}_{\rm sys}]- \int d\omega \kappa_i(\omega) [\hat{\sigma}^-_i(t),\hat{\sigma}^+_i(t)]\hat{b}(t,\omega)\nonumber \\
\!\!\!\!\!\!&\rightarrow& -i\omega \hat{\sigma}^-_i(t)-i\sum_{j} g_{ij} \hat{\sigma}^-_j (t)+ \int d\omega \kappa_i(\omega) \hat{\sigma}_{zi}(t)\hat{b}(t,\omega)\nonumber
\eea where in the second line we have made use of the commutator relation $[\hat{\sigma}_i^+(t),\,\hat{\sigma}_j^-(t)]=\hat{\sigma}_{zi}(t)$, where $\hat{\sigma}_{zi}(t)$ is the time-dependent Heisenberg picture Pauli-Z spin operator for the $i$th mode. 

We can proceed with the input-output treatment performed in Section II. For simplicity, we will consider the case of uncoupled fermionic qubits ($g_{ij}=0$). Analogously to the last line of (\ref{HFerm}) describing the lowering operator, we find that the Pauli-Z spin operators obey a Heisenberg equation of motion

\bea\label{HFermZ}
\dot{\hat{\sigma}}_{zi}(t)&=& -\int d\omega \kappa_i(\omega) \left(\hbd(\omega)\hat{\sigma}^-_i(t) +\hat{\sigma}^+_i(t) \hb(\omega) \right). \nonumber\\
\eea 

Now, we formally solve for the evolution of the continuum operator $\dot{\hat{b}}(t,\omega)$ as in (\ref{formsol}), invoke the first Markov approximation ($\kappa_i(\omega) = \sqrt{\frac{\gamma_i}{2\pi}}$), and define an input field operator as in (\ref{bin}) to result in equations of motion with the continuum field integrated out

\begin{widetext}\bea\label{quantumlangevinfermion}
\phantom{blah}\dot{\hat{\sigma}}^-_i(t)&=& -i\omega_i \hat{\sigma}^-_i(t)  + \sum_j \frac{\sqrt{\gamma_i\gamma_j}}{2} \hat{\sigma}_{zi}(t) \hat{\sigma}^-_j(t)+ \sqrt{\gamma_i} \hat{\sigma}_{zi}(t)  \hat{b}_{\rm in}(t) \nonumber \\
\phantom{blah}\dot{\hat{\sigma}}_{zi}(t)&=& - \sqrt{\gamma_i} \left(  \hat{b}_{\rm in}^+(t)\hat{\sigma}^-_i(t) + \hat{\sigma}^+_i(t) \hat{b}_{\rm in}(t)\right) 
+ \sum_j \frac{\sqrt{\gamma_i\gamma_j}}{2} \left(\hat{\sigma}^+_j(t)\hat{\sigma}^-_i(t) + \hat{\sigma}^+_i(t)\hat{\sigma}^-_j(t)\right).
\eea\end{widetext}

Notably, even in the zero-temperature limit where $\hat{b}_{\rm in}(t) \rightarrow 0$, the equations are nonlinear: lower order operator evolution will depend on higher order operator evolution. In general, there are two solutions. The first is to truncate this dependence, and set some higher order expectation value to zero (or to an experimentally measured value). The second is to take the low-intensity few-photon limit, such that $ \hat{\sigma}_{zi}(t) \rightarrow -1$, recovering the bosonic quantum Langevin equation derived in Section II. This is the approach taken here.

}

\section{Gramm matrix for two uncoupled bosonic modes at zero temperature}

We now go through the process of solving for the Gramm matrix (also known as the dynamical matrix) for two collectively decaying, uncoupled bosonic modes at zero temperature, first for the state amplitudes and then for the number operator. 

We begin by writing the generalized Langevin equation (\ref{quantumlangevin}) in vector form

\bea\label{vectorlangevin}
\frac{d}{dt}\vec{c}(t) = M \vec{c}(t) - \vec{S}
\eea where $\vec{c}(t)$ is the time-dependent vector of expectation values $\expect{\hat{c}_i(t)}$, and $\vec{S}$ is a vector of expectation values $\sqrt{\gamma_i} \expect{\hat{b}_{\rm in}(t)}$, which is zero in the absence of a coherent drive. For two uncoupled bosonic modes, the Gramm matrix $M$ has the form 

\bea\label{2gramm}
M = \begin{bmatrix}
-i\omega_1 -\frac{\gamma_1}{2} & -\frac{\sqrt{\gamma_1\gamma_2}}{2} \\
-\frac{\sqrt{\gamma_1\gamma_2}}{2} & -i\omega_2 -\frac{\gamma_2}{2}
\end{bmatrix}
\eea with eigenvalues 

\begin{widetext}
\bea\label{eigenvals}
\lambda_\pm = \frac{-i (\omega_1+\omega_2) + \frac{\gamma_1+\gamma_2}{2} \pm \sqrt{\left(-i (\omega_1+\omega_2) + \frac{\gamma_1+\gamma_2}{2}\right)^2-4\left(i(\frac{\omega_1\gamma_2}{2} + \frac{\omega_2 \gamma_1}{2}) - \omega_1\omega_2 - \frac{\gamma_1\gamma_2}{4}\right)}}{2}
\eea\end{widetext} each associated with an eigenvector $\vec{v}_{\pm}$, omitted here for brevity.

The solution for the evolution of the components of $\vec{c}(t)$, which we denote $c_i(t) = \expect{\hat{c}_i(t)}$, are then given 

\bea\label{csolAmps}
c_1(t) &=& A_1 e^{\lambda_+ t} v_{+,1} + A_2 e^{\lambda_- t} v_{-,1} \nonumber \\
c_2(t) &=& A_1 e^{\lambda_+ t} v_{+,2} + A_2 e^{\lambda_- t} v_{-,2} 
\eea where $v_{\pm,i}$ is the $i$th component of the eigenvector $\vec{v}_{\pm}$, and the coefficients $A_1$ and $A_2$ are determined by initial conditions numerically. 

Similarly, we can write a vector $\vec{N}$ whose components are $\expect{\hat{N}_1}$, $\expect{\hat{N}_2}$, $\expect{\hat{O}_{12}}$, and $\expect{\hat{Y}_{12}}$. The evolution of $\vec{V}$, as defined by (\ref{NumberEvolution}-\ref{auxEqs|:1}), is then described by the matrix equation

\bea\label{vectorlangevin4}
\frac{d}{dt}\vec{N}(t) = M' \vec{N}(t) - \vec{S}'
\eea where again we have defined a source vector $\vec{S}'$ (which is zero for zero-temperature and no coherence in the field) and a Gramm matrix

\bea\label{4gramm}
M' = \begin{bmatrix}
-\gamma_1 & 0 & -\frac{\sqrt{\gamma_1\gamma_2}}{2} &0 \\
0&-\gamma_2  & -\frac{\sqrt{\gamma_1\gamma_2}}{2} &0 \\
-\sqrt{\gamma_1\gamma_2}&-\sqrt{\gamma_1\gamma_2} &-\frac{\sqrt{\gamma_1\gamma_2}}{2}&i(\omega_1 - \omega_2) \\
0 & 0  &i(\omega_1 - \omega_2)&-\frac{\sqrt{\gamma_1\gamma_2}}{2} \\
\end{bmatrix}.
\eea

As before, we can solve for the dynamics of the vector components of $\vec{N}$ (that is $N_1\equiv\expect{\hat{N}_1}$, $N_2\equiv\expect{\hat{N}_2}$, $N_3 \equiv \expect{\hat{O}_{12}}$, and $N_4 \equiv \expect{\hat{Y}_{12}}$) by writing them in terms of the four eigenvalues of $M'$ denoted $\lambda_i'$ as well as the components of the four eigenvectors $\vec{v}_i'$ of $M'$, which we denoted $v_{i,j}$ with the first index labeling the $i$th eigenvector and the second index labeling its $j$th component. This results in solutions 

\bea\label{csolAmps}
N_1(t) &=& A_1 e^{\lambda_1 t} v_{1,1}' + A_2 e^{\lambda_2 t} v_{2,1}' + A_3 e^{\lambda_3 t} v_{3,1}'\nonumber  + A_4 e^{\lambda_4 t} v_{4,1}' \\
N_2(t) &=& A_1 e^{\lambda_1 t} v_{1,2}' + A_2 e^{\lambda_2 t} v_{2,2}' + A_3 e^{\lambda_3 t} v_{3,2}'\nonumber  + A_4 e^{\lambda_4 t} v_{4,2}' \\
N_3(t) &=& A_1 e^{\lambda_1 t} v_{1,3}' + A_2 e^{\lambda_2 t} v_{2,3}' + A_3 e^{\lambda_3 t} v_{3,3}'\nonumber  + A_4 e^{\lambda_4 t} v_{4,3}' \\
N_4(t) &=& A_1 e^{\lambda_1 t} v_{1,4}' + A_2 e^{\lambda_2 t} v_{2,4}' + A_3 e^{\lambda_3 t} v_{3,4}'\nonumber  + A_4 e^{\lambda_4 t} v_{4,4}' \\
\eea with the coefficients $A_i$ determined from initial conditions, as before.

\end{document}